\documentclass[aps,prb,twocolumn,floats]{revtex4}

\usepackage{amssymb}
\usepackage{amsmath}
\usepackage{graphicx}
\usepackage{subfigure}
\usepackage{bm}
\usepackage{mdframed}
\usepackage[pangram]{blindtext}
\usepackage{bbm}

\begin{document}

\title{Double single-channel Kondo coupling in graphene with Fe molecules}

\author{I. M. Vicent,$^1$ L. Chirolli,$^{2,3}$ and F. Guinea,$^{1,4}$}

{\affiliation{$^1$Instituto Madrile\~no de Estudios Avanzados en Nanociencia (IMDEA-Nanociencia), 28049 Madrid, Spain\\
$^2$Department of Physics, University of California, Berkeley, California 94720, USA\\
$^3$Instituto Nanoscienze-CNR, I-56127 Pisa, Italy\\
$^4$Donostia International Physics Center (DIPC) UPV/EHU, E-20018, San Sebasti\'an, Spain}

\begin{abstract}
We study the interaction between graphene and a single-molecule-magnet, $[Fe_4(L)_2(dpm)_6]$. Focusing on the closest Iron ion in a hollow position with respect to the graphene sheet, we derive a channel selective tunneling Hamiltonian, that couples different $d$ orbitals of the Iron atom to precise independent combinations of  sublattice and valley degrees of freedom of the electrons in graphene. When looking at the spin-spin interaction between the molecule and the graphene electrons, close to the Dirac point the channel selectivity results in a channel decoupling of the Kondo interaction, with  two almost independent Kondo systems weakly interacting among themselves. The formation of magnetic moments and the development of a full Kondo effect depends on the charge state of the graphene layer.

\end{abstract}
 
\maketitle

\textit{Introduction.---}
Since the low-temperature magnetic phase transition of metals with magnetic impurities diluted in was explained by Kondo\cite{Kondo} in the middle of the past century, the theoretical framework has been extensively developed.\cite{SchriefferWolff, CoqblinBlandin,CoqblinSchrieffer,Anderson,Wilson} Over time, the problem has been extended, particularly with the discovery of the multichannel Kondo effect\cite{Muramatsu} and the study of the interactions between two magnetic impurities.\cite{Varma1,Varma2} In carbon 2D materials, a double degeneracy appears in spin and orbital momentum. This double degeneracy allows the emergence of a symmetric SU(4) Kondo effect with a strong coupling between the spin and orbital degrees of freedom.\cite{Jarillo-Herrero, Ramon, Goldhaber, Weymann}
The special condition of graphene, positioned halfway between metals and semiconductors, appears as an appealing scenario for Kondo physics. Its linear dispersion relation, and the ease to tune its chemical potential are important elements too.\cite{Fradkin1,Ogata,pseudogapKondo,Uchoa2} Atomic vacancies in the graphene lattices were the first system in which magnetic transitions were founded.\cite{Brihuega} The relaxation of the lattice around the vacancy can be solved changing the vacancy by an hydrogen impurity localized on top site.\cite{Hector} The strong coupling of the impurity and the carbon atom generates a localized magnetic moment by the subtraction of one electron from the Fermi sea.\cite{EvaAndrei}
Magnetic molecules\cite{molecula1} offer an easy way to study the magnetic interaction between graphene and localized spin moments due to its clear magnetic properties. Functionalized hybrid of graphene and one of this molecules, $[\mathrm{Fe}_4(L)_2(dpm)_6]$, shows interesting properties for the study of the magnetic phase transition.\cite{LapoNat}
  
  In this work we describe in detail the interactions between a magnetic molecule and graphene.  The interplay between the different degrees of freedom of the graphene electrons (valley, pseudospin, and spin) with the projection of the angular momentum and the spin of the electrons in the molecule leads to a channel selectivity of the coupling. The type and number of the channels that can tunnel to the molecule depends on the site that the molecule occupies on the lattice. If the molecule is in the center of the hexagon just two independent combinations of valley, pseudospin and spin  lives in. Each one and each subspace are tunnel coupled to the two relevant orbitals of the Fe(III) ion in the core of the molecule, giving rise to a double single-channel Kondo effect, as we observe the absence of the Ruderman-Kittel-Kasuya-Yoshida (RKKY) interaction between the two orbitals of the molecule, due to the independence of the electronic channels involved. \begin{figure}
\includegraphics[width=.99\columnwidth]{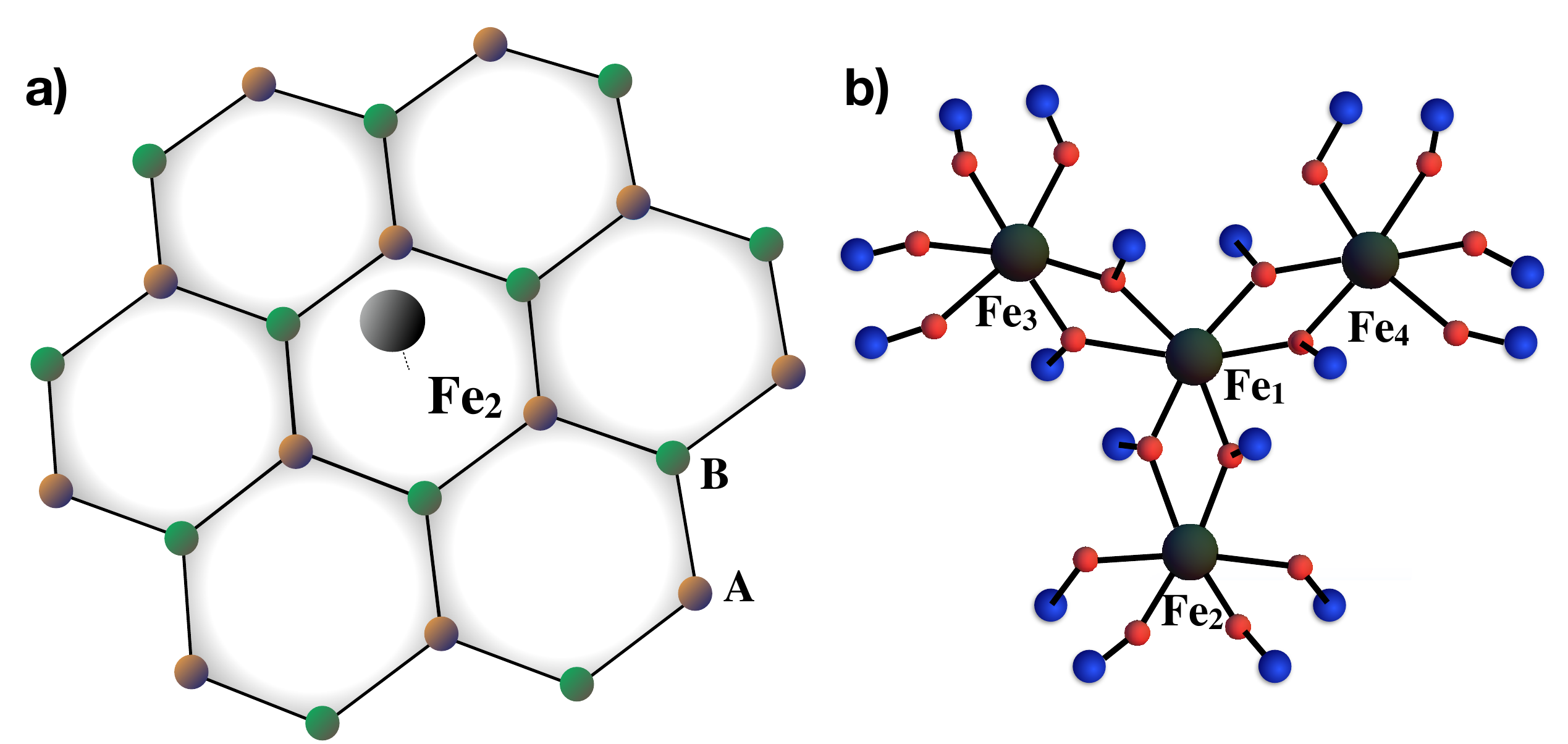}
\caption{\textbf{a}, Schematic representation of an impurity on hollow position, at the center of the graphene hexagon. In our case we model the molecule by considering only the nearest Fe ion (Fe$_2$). \textbf{b}, Scheme of the molecular core of the $[\mathrm{Fe}_4(L)_2(dpm)_6]$. The Fe ions (black) are coupled via pairs of oxygen atoms (red). Hydrogen atoms and most external carbon atoms have been omitted for clarity.}
\label{fig:Impurity}
\end{figure}\\

\textit{The model.---}
We consider a graphene sheet described by a nearest neighbor hopping Hamiltonian 
\begin{equation}
H_G=-t\sum_{\left<i,j\right>,s}a^\dagger_{s,i}b_{s,j}+\mathrm{h. c.},
\end{equation}
where $a,b$ are fermionic operators that annihilate an electron with spin $s=\uparrow\downarrow$, on sublatices A and B, respectively, and $t\sim2.7$ eV is the hopping energy between next nearest neighbors. In momentum space hopping between $a_{\bf k}$ and $b_{\bf k}$ is specified by
$\gamma_{\bf k}=-t\sum_{i}e^{i\mathbf{k}\cdot\mathbf{c}_j}$, with $\mathbf{c}_1=\hat{x}/\sqrt{3}$, $\mathbf{c}_2=-(\hat{x}+\sqrt{3}\hat{y})/(2\sqrt{3})$, $\mathbf{c}_3=(-\hat{x}+\sqrt{3}\hat{y})/(2\sqrt{3})$, nearest-neighbors vectors. Expansion of $\gamma_{\bf k}$ around the Dirac points $\pm {\bf K}=(0,\pm 2\pi/3)$ produces the celebrated graphene Dirac Hamiltonian.

We then consider a single-molecule-magnet (SMM), $[Fe_4(L)_2(dpm)_6]$ \cite{molecula1, LapoNat} added on top of the graphene sheet. It is a molecule with a $S=5$ spin moment due to the four Fe(III) ions in its core\cite{molecula2}, and it is well described by the Hamiltonian \cite{LapoNat}
 \begin{equation}
 H_{SSM}=\sum_{i=2}^4 J_{mol}\left({\mathbf S_1} \cdot {\mathbf S_i}\right)+\sum_{\langle i, j\rangle (i,j \neq 1)} J_{mol}'\left({\mathbf S_i} \cdot {\mathbf S_j}\right),
 \end{equation} 
where $J_{mol}$ and $J'_{mol}$ are the exchange couplings between nearest and next nearest Iron atoms, respectively. $\mathbf{S}_i$ are the $S=5/2$ spin matrices describing the spin of each iron ion. As shown in Fig.\ref{fig:Impurity}, $\mathrm{Fe_1}$ is one ion surrounded by the other three. The coupling $J_{mol}>0$ is antiferromagnetic whereas the coupling $J_{mol}'<0$ is ferromagnetic, with $|J'_{mol}|\ll J_{mol}$.  This way, the three outer ions, Fe$_{2,3,4}$ anti align with respect to the central ion Fe$_1$. 

Since the distance of the graphene sheet to the ion Fe$_2$ is more than twice smaller than the distance to the other ions, the tunneling from graphene to the molecule involves only the ion Fe$_2$, see Fig. \ref{fig:Impurity}. We then focus on Fe$_2$ and consider the effect of other three ions by their contribution in the energy levels of Fe$_2$. Fe(III) has five electrons distributed in five spin-degenerate outer most $d$-orbitals. In the $S=5/2$ state, all five $d$-orbitals are singly occupied
and the spin of the ion can be described as the sum of five localized spin 1/2. The Hamiltonian of the ion Fe$_2$ reads
 \begin{equation}
 H_{\rm Fe_2}=\sum_{m,s}\epsilon_{m,s}d^{\dagger}_{m,s}d_{m,s}+H_U,
 \end{equation} 
where $d_{m,s}$ are fermionic operators describing the five $d$-orbitals, $\epsilon_{m,\sigma}$ their associated energies, $m$ is the angular moment projection of the different states of angular momentum $l=2$. The symmetry of the system, $(C_{3v})$, shifts the energy of the orbitals of the Fe(III) ion creating two pairs of degenerate states with the same $|m|\neq 0$. In the case of the Fe (III) ions the lowest levels are the doublets $d_{\pm}=(d_{x^2-y^2}\pm i d_{xy})\sqrt{2}$.

The electrostatic repulsion between the Fe(III) electrons is measured by $U_{mm'}$, and the exchange energy $J_{mm'}$, between the localized electrons in different orbitals.
 \begin{equation} 
 \begin{split}H_U&=\sum_{m,m'}U_{m,m'}\hat{n}_{m,\uparrow}\hat{n}_{m',\downarrow}
 \\+&\frac{1}{2}\sum_{s}\sum_{m\not=m'} \left(U_{m,m'}-J_{m,m'}\right)\hat{n}_{m,s}\hat{n}_{m',s}.
 \end{split}
 \end{equation}
where $\hat{n}_{m,s}=d^\dag_{m,s}d_{m,s}$. For simplicity we are going to consider $U_{mm}=U_{mm'}$ and $J_{mm}=J_{mm'}$.\\

{\it Tunneling Hamiltonian.---}
The most stable position for atoms such as Fe on a graphene lattice is at the center of the hexagons, the hollow site\cite{hollow-top1,hollow-top2,hollow-top3}, which is the case that we will consider here. For the hollow position the Hamiltonian describing the tunneling between the localized states of the impurity and the conduction electrons of graphene can be generically written as  
\begin{equation}
H_V=\sum_{m,s}\sum_{i=1}^3 \left[V^m_{a,i}a^{\dagger}_s(\mathbf{a}_i)+V^m_{a,i}b^{\dagger}_s(-\mathbf{a}_i) \right]d_{m,s}+ \mathrm{h. c.}, 
\end{equation}
where $V^m_{c,i}$ $(c=a,b)$ are tunneling matrix elements between the localized states with angular momentum $m$ with each of the six carbon atoms surrounding the hollow position.  These can be specified by a unique Slater-Koster-like matrix element $V$. States with $m=\pm 2$ on the Fe ion are invariant under $C^z_\infty$, and that the tunneling process selects  combinations of $a_{\bf k}$ and $b_{\bf k}$ that are $C^z_3$ invariant. This yields a selectivity between angular momentum $m$, sublattice and valley.\cite{anex1} By expanding the expression around the Dirac points the tunneling Hamiltonian takes the form,
\begin{equation}\label{HV}
\begin{split}
H_V &= \frac{V}{\sqrt{2}}\sum_{{\bf k}}d_{+}^\dagger (a_{\mathbf{k},K}+b_{\mathbf{k},K'})
+d_{-}^\dagger (a_{\mathbf{k},K'}+b_{\mathbf{k},K})
+{\rm H.c.},
\end{split}
\end{equation}
where a factor $3/\sqrt{2}$ has been reabsorbed in $V$ and the spin label has been suppressed, as the tunneling conserves the spin. 
This expression can be considered as a low energy expansion around the Dirac point of graphene, in powers of the electron energy, $\epsilon / W$, where $W$ is the width of the $p_z$ band in graphene. Corrections to the couplings, as the momenta of the graphene states deviates from the Dirac point are neglected.

\begin{figure}
\includegraphics[width=.8\columnwidth]{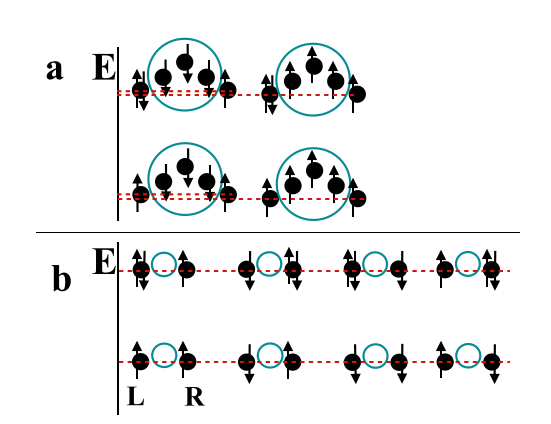}
\caption{ Diagram of the energies of the different states considered for the Fe(III) ion, including the splitting due to ligand fields due to the environment. \textbf{a}, The energy splitting due to the spin flip of the $|m|=2$ electrons is negligible. We consider the other three orbitals as frozen and as their $S_z$ is not relevant for the energy of the system we neglect it. 
\textbf{b}, The system can be described by the eight states shown in the figure. Double occupied states lie at higher energies. 
}
\label{fig:scheme}
\end{figure}

The tunneling term connects the graphene states only with $|m|=2$ orbitals of Fe$_2$. The ground state confi-guration of the Fe ions consists in five electrons with the same $S_z=\pm 1/2$, the ground state has $S=5/2$, and the first excited states with $S=3/2$, $1/2$ are 7 and 11 meV above the ground state\cite{molecula1}. The states with integer $S_z$ are much more energetic, as schematically depicted in Fig.~\ref{fig:scheme}. Due to the small difference between the four low-energy states, we can neglect the splitting between them, and consider all of them as degenerate. This assumption implies that the energy required to flip the spin of a given orbital is negligible. We keep, on the other hand, the crystal field splitting between orbitals with different values of $|m|$, of order $1470 \, \mathrm{cm}^{-1} \approx 0.18 \, \mathrm{eV}$\cite{molecula1}. Hence, the interaction between the graphene electrons and the Fe$_2$ ion is through the atomic orbitals of energy closest to the Dirac point of graphene, which we assume to have $m = \pm 2$ (the calculations are equivalent for $m = \pm 1$).

We now diagonalize the graphene Hamiltonian around the Dirac points and introduce eigenoperators $c_{\mathbf{k},\beta,\tau,s}=\left(a_{\mathbf{k},\tau,s}+\beta\tau e^{i\tau\theta_\mathbf{k}} b_{\mathbf{k},\tau,s} \right)/ \sqrt{2}$,  with $\tan\theta_\mathbf{k}=k_y/k_x$, $\beta=\pm$ distinguishing between valence and conduction bands, and $\tau=\pm$ indexing the valley. The Hamiltonian of the Fe$_2$ ion and the graphene electrons reads
\begin{eqnarray}
H_0&=&\sum_{\mathbf{k},s,\tau,\beta}\epsilon_{\beta,\mathbf{k}} c^{\dagger}_{\mathbf{k},\beta,\tau,s}c_{\mathbf{k},\beta,\tau,s}+\sum_{m=\pm2,s}\epsilon_0 d_{m,s}^{\dagger}d_{m,s}\nonumber\\
&+&\frac{U}{2} N_{d} \left(N_d-1\right).
\end{eqnarray}
The form of the tunneling suggests the introduction of two new fermionic operators representing the two independent combinations of sublattice and valley appearing in Eq.~(\ref{HV}), $C_{{\bf k},\alpha,s}=\sum_{\beta,\tau} A^\alpha_{\beta,\tau}c_{{\bf k},\beta,\tau,s}$. The tunneling term in the new basis reads $H_V=V\sum_{\mathbf{k}, \alpha,s}  C^\dag_{\mathbf{k},\alpha,s} d_{\alpha,s}+{\rm H.c.}$, where the two $m$ orbitals have been relabelled as $\alpha=L,R=\pm 1$\cite{anex1}, in analogy with a double-dot configuration.\\

{\it Effective low-energy model.---} 
In order to see the conditions for which the hopping to graphene can quench this localized moment, we derive an effective spin-spin coupling between the graphene electrons and the Fe$_2$ ion via eliminating the tunneling at first order through the well known Schrieffer-Wolff  transformation\cite{SW}. The latter consists in defining a new Hamiltonian $\tilde{H}$ that is obtained via a unitary transformation 
$
\tilde{H}=e^{S}He^{-S}=H+[S,H]+\frac{1}{2}[S,[S,H]]+\ldots, 
$
with $S$ an anti Hermitian operator on order of the tunneling $H_V$. By requiring 
\begin{equation}
\left[H_0,S\right]=H_V,
\end{equation} 
the tunneling Hamiltonian $H_V$ is eliminated at first order. By further truncating the expansion at second order the effective Hamiltonian reads
\begin{equation}
\tilde{H}=H_0+\frac{1}{2}[S,V].
\end{equation}
In this problem the existence of different interacting electrons in the ion adds a non-trivial difficulty. 
Given the form of $H_V$, the operator $S$ is given by the following expression
\begin{equation}
S=\sum_{ij,\alpha,\mu,\tau,\mathbf{k},s}\frac{ V^\alpha_{\mu\tau}P_id^\dag_{\alpha,s}c_{\mu,\tau,\mathbf{k},s}P_j}{\epsilon_0+(i+j-1)U-\epsilon_{\mu,\mathbf{k}}}-{\rm H.c.}.
\end{equation}
where $V^\alpha_{\mu\tau}=VA^\alpha_{\mu \tau}$ and $P_i$ are the projector operators for the double dot system that satisfy $\sum_{i=0,4} P_i=1$, with $i$ labelling the number of electrons in the double dot. The correction to the unperturbed Hamiltonian $H_0$ is composed by several terms and takes the form 
\begin{equation}
H'=\frac{1}{2}[S,H_V]=H_\mathrm{K}+H_{ch}+H_{mix}+H_C.
\end{equation} 
The result consists in a Kondo term and a charge term, plus a mixing term and a Cooper term. The latter is composed by terms like $c^\dag c^\dag d d$, that we discard. The Hamiltonian $H'$ needs to be projected onto the desired subspace, that for the present problem is the double occupancy subspace.  Therefore, a further step is carried on as $H'\to P_2 H'P_2$. 

\textit{Effective Kondo Model.---} 
The double occupancy subspace is in turn composed by states with one electron per dot and states with two electrons in one dot and zero in the other.  However if we consider instead of  the equal repulsion interaction for the intra-orbital and inter-orbital case that one in which, $U_{m,m} \gg U_{m,m'}$ with $m'\neq m$, we can neglect the double occupancy cases. We then carry on a second projection as $H'\to P_{1L}P_{1R}H'P_{1L}P_{1R}=H_{\rm K}+H_{ch}$. Introducing a cumulative index $\mu\equiv\{\mu,\tau,{\bf k}\}$, the full effective Kondo Hamiltonian reads
\begin{eqnarray}
H_{\rm K}&=&-\sum_{\alpha\mu\mu'} \frac{J^\alpha_{\mu,\mu'}}{2}\left[c^\dag_{\mu'\downarrow}c_{\mu\uparrow}d^\dag_{\alpha\uparrow}d_{\alpha \downarrow}+c^\dag_{\mu'\uparrow}c_{\mu\downarrow}d^\dag_{\alpha\downarrow}d_{\alpha \uparrow}\right.\nonumber\\
&+&\frac{1}{2}\left.(c^\dag_{\mu'\uparrow}c_{\mu \uparrow}-c^\dag_{\mu'\downarrow}c_{\mu \downarrow})(d^\dag_{\alpha\uparrow}d_{\alpha \uparrow}-d^\dag_{\alpha\downarrow}d_{\alpha \downarrow})\right],
\end{eqnarray}
with the coupling constant given by
\begin{eqnarray}
J^\alpha_{\mu,\mu'}&=&V^\alpha_\mu(V^\alpha_{\mu'})^*\left[\frac{1}{\epsilon_0+2U-\epsilon_\mu}+\frac{1}{\epsilon_0+2U-\epsilon_{\mu'}}\right]\nonumber\\
&-&V^\alpha_\mu(V^\alpha_{\mu'})^*\left[\frac{1}{\epsilon_0+4U-\epsilon_\mu}+\frac{1}{\epsilon_0+4U-\epsilon_{\mu'}}\right].~~~~
\end{eqnarray}

The Kondo Hamiltonian is still quite involved at this stage, as it mixes different bands and couples different channels. However, close to the Dirac point we have that $\epsilon_{\beta,\mathrm{k}} \ll \epsilon_0 , U$ and we can safely neglect the energy dependence in the coupling. This procedure highly simplifies the expression and highlights the channel selectivity contained in the tunneling Hamiltonian $H_V$. Due to destructive interference the two orbitals $L$ and $R$ separately couple to two independent channels in the valley and sublattice space. 

At the Dirac points at $K$ and $K'$ we define the graphene eight-component spinor  $\Psi_{\bf k}$
and the iron 2-component spinor $\Psi_{\alpha}^\dagger=(d_{\alpha,\uparrow}^\dagger,d_{\alpha,\downarrow}^\dagger)$. The expression for the Kondo Hamiltonian reads
\begin{equation}
H_K=-J\sum_{\mathbf{k,k}',\alpha}\Psi^\dagger_{\bf k} \Sigma_\alpha\mathbf {s}\Psi_{{\bf k}'} \cdot \Psi_\alpha^\dagger{\mathbf S}_\alpha\Psi_\alpha,
\label{kondo}
\end{equation}
where $\textbf{s}$ and $\bf{S_\alpha}$ are the spin 1/2 Pauli matrix vectors of graphene and of the different orbitals $(\alpha=L,R)$ of the iron.\cite{anex1} The operators $\Sigma_{\alpha=L,R}$ are given by
\begin{eqnarray}
\Sigma_R&=&1-\sigma_z\tau_z+\sigma_x\tau_x+\sigma_y\tau_y,\\
\Sigma_L&=&1+\sigma_z\tau_z+\sigma_x\tau_x-\sigma_y\tau_y,
\end{eqnarray}
with $\sigma_i$, $\tau_i$, sublattice and valley Pauli matrices, and define two independent channels in sublattice and valley space, $\Sigma_L\Sigma_R=0$. We then see that close to the Dirac point a channel decoupling takes place and the two iron orbitals couple to different subspaces of the graphene Hamiltonian, yielding a double single-channel Kondo Hamiltonian. The coupling constant $J$ reads
\begin{equation}
J=\frac{4V^2}{(\epsilon_0+2U)(\epsilon_0+4U)},
\end{equation}
in complete analogy with the original Kondo Hamiltonian derived by Schrieffer and Wolff from the single impurity Anderson model\cite{SW}. The sign of the coupling is negative for $\epsilon_0+2U<0$, that is assumed to be the ground state energy of the doubly occupied double dot problem, so that the overall Kondo spin-spin interaction is antiferromagnetic.

In addition to the Kondo term, Eq.(\ref{kondo}), the spin independent part of the coupling can be written as
\begin{equation}\begin{split}
H_{ch}&=-\sum_{\bf{k,k}',\alpha=\pm}\left[W+\frac{J}{4}\Psi_\alpha^\dagger\Psi_\alpha\right] \Psi^\dag_{\bf k}(1+\alpha \sigma_z\tau_z)\Psi_{{\bf k}'},
\end{split}
\end{equation}
with $W=V^2/(\epsilon_0+4U)$ and $\alpha=\pm$ applying to $L$ and $R$, respectively. This term does not affect the spin of the itinerant electrons in graphene. 

For completeness, we note that there is a third term,  $H_{mix}$, which takes into account the mixing between the two channels. This term arise when projecting on the subspace with double-occupancy in one orbital and zero occupancy on the other and is suppressed,
\begin{equation}
H_{mix}=\sum_{\mathbf{k},{\bf k}'}J'\left(\alpha_x(\sigma_x+\tau_x)+\alpha_y(\sigma_y\tau_z+\sigma_z\tau_y)\right),
\end{equation}
where $\alpha_+=d^\dag_{L}d_R$, $\alpha_-=d^\dag_{R}d_L$ and $\alpha_\pm=(\alpha_x\pm\alpha_y)/2$ and the coupling at the Dirac point is given by
\begin{equation}
J'=V^2\left[\frac{-1}{\epsilon_0+4U}+\frac{1}{\epsilon_0+2U}\right].
\end{equation}
One can see how this term does not include spin-flip processes. 

{\it RKKY interaction.---}
Finally, we look at the possibility that the graphene electrons mediate an effective interaction between the $L$ and $R$ spins through the RKKY interaction. Considering that the two spins sit at the same spatial position, the RKKY interaction at second order in the coupling $J$ reads 
\begin{equation}
H_{\mathrm{RKKY}}=\sum_{\mu,\nu}\chi_{\mu,\nu}\hat{S}^L_\mu\hat{S}^R_\nu,
\end{equation}
with $\mu,\nu=x,y,z$ labelling the spin components and with the effective spin-spin susceptibility $\chi_{\mu,\nu}$ defined as
\begin{equation}
\chi_{\mu,\nu}=\sum_{k, k'}J^2\frac{1}{\beta}\sum_{i \omega_n}\mathrm{Tr} \left[\Sigma_L s_\nu G^0_{\omega_n,k}\Sigma_R s_\mu G^0_{\omega_n,k'}\right].
\end{equation}
Having neglected the dependence of $J$ on the momentum, integration of the graphene Green's function over momentum rules out its matrix structure and we are left with
\begin{equation}
\chi_{\mu,\nu}\propto \mathrm{Tr}[s_\mu \Sigma_L s_\nu \Sigma_R]=0,
\end{equation}
where the last equality follows from the orthogonality of the two channels defined by $\Sigma_{L,R}$. We then conclude that the model Eq.~(\ref{kondo}) effectively described two independent single-channel Kondo Hamiltonians. \\

\textit{Discussion}.---
To summarize, we have studied the interaction between graphene and a molecular magnet. We have focused in the most stable and symmetric case, with the core of the molecule being at the center of a graphene hexagon. The leading coupling, Eq.~(\ref{kondo}), shows two Kondo couplings involving the $m = \pm 2$ orbitals in the iron ion nearest to the graphene layer. These two Kondo systems involve different combinations of valley and sublattice indices in the graphene layer. Exactly at the Dirac point, the vanishing density of states of graphene makes the Kondo coupling irrelevant, although two Kondo singlets will be formed for sufficiently large values of the coupling $J$. Away from the Dirac point, the Kondo coupling becomes marginally relevant, and, at the same time, an effective coupling between the two Kondo systems will develop through the graphene electrons. These two effects are of comparable strength, and they can lead to a rich phase diagram. The nature of the most stable phases depends on details at the atomic scale outside the scope of this work.

\section*{Acknowledgements}
This work was supported by funding from the European Commission, under the Graphene Flagship, Core 3, grant number 881603, and by the grants NMAT2D (Comunidad de Madrid, Spain),  SprQuMat and SEV-2016-0686, (Ministerio de Ciencia e Innovaci\'on, Spain). L.C. also acknowledges the European Commission for funding through the MCSA Global Fellowship grant TOPOCIRCUS-841894.

\section{Supplementary material}

\subsection{Tunneling Hamiltonian}
We consider orbitals $d_{x^2-y^2}$ and $d_{xy}$ in the hollow position and allow a tunneling matrix element $V$ with the underlaying $p_z$ orbitals on the hexagon. The Hamiltonian reads
\begin{eqnarray}
H_V&=&Vd^\dag_{x^2-y^2}\sum_{\alpha=a,b}\left[\alpha_1+\alpha_2\cos\frac{\pi}{3}+\alpha_3\cos\frac{2\pi}{3}\right]\nonumber\\
&+&Vd^\dag_{xy}\sum_{\alpha=a,b}\left[\alpha_2\sin\frac{\pi}{3}+\alpha_3\sin\frac{2\pi}{3}\right]+{\rm H.c.}
\end{eqnarray}
Introducing $d_\pm=(d_{x^2-y^2}\pm i d_{xy})/\sqrt{2}$ we have
\begin{equation}
H_V=\frac{1}{\sqrt 2}d^\dag_\pm\sum_{\alpha=a,b}(\alpha_1+e^{\mp 2\pi i/3}\alpha_2+e^{\pm 2\pi i/3}\alpha_3)\rm H.c.+{}
\end{equation}
In momentum space around the Dirac point we have
\begin{eqnarray}
H_V&=&\frac{1}{\sqrt 2}d^\dag_\pm a_{{\bf k},\tau}(1+e^{2\pi i(\tau\mp 1)/3}+e^{-2\pi i(\tau\mp 1)/3})\nonumber\\
&+&\frac{1}{\sqrt 2}d^\dag_\pm b_{{\bf k},\tau}(1+e^{2\pi i(\tau\pm 1)/3}+e^{-2\pi i(\tau\pm 1)/3})+{\rm H.c.}\nonumber\\
\end{eqnarray}
so that the orbital valley and sublattice selectivity appears as an interference effect. We can define two independent fermionic operators
\begin{eqnarray}\label{EqC2}
C_{\mathbf{k},L}&=&\frac{1}{\sqrt 2}\left[a_{\mathbf{k},K}+b_{\mathbf{k},K'}\right],\\
C_{\mathbf{k},R}&=&\frac{1}{\sqrt 2}\left[c_{\mathbf{k},K'}+b_{\mathbf{k},K}\right].
\end{eqnarray}

\subsection{Schrieffer-Wolff transformation}

We look for an operator $S$ that satisfies
\begin{equation}
[H_0,S]=H_V
\end{equation}
We introduce a complete set of projector operators $P_i$ (satisfying $P_i^\dagger=P_i$ and $P_i^2=P_i$) for the double dot system that satisfies $\sum_{i=0,4}P_i=1$. Explicitly they are given by
\begin{equation}
\begin{array}{lll}
P_0&=&P_{0L}P_{0R},\\
P_1&=&P_{1L}P_{0R}+P_{0L}P_{1R},\\
P_2&=&P_{2L}P_{0R}+P_{1L}P_{1R}+P_{0L}P_{2R},\\
P_3&=&P_{2L}P_{1R}+P_{1L}P_{2R},\\
P_4&=&P_{2L}P_{2R},
\end{array}
\end{equation}
where $P_{1,\alpha}=n_{\alpha\uparrow}(1-n_{\alpha,\downarrow})+n_{\alpha\downarrow}(1-n_{\alpha,\uparrow})$ and so on (we need all of them). 

We write the tunneling Hamiltonian as
\begin{equation}
H_V=\sum_{\alpha,\gamma,s}V^\alpha_{\gamma}d^\dag_{\alpha,s}c_{\gamma,s}+{\rm H.c.}\nonumber
\end{equation}
where $\gamma\equiv\{\beta,\tau,{\bf k}\}$ is a cumulative label that includes the band label, the valley and the momentum. 

Given the form of $H_V$, the operator $S$ is given by the following expression
\begin{equation}
S=\sum_{ij,\alpha,\gamma,s}V^\alpha_{\gamma}\frac{P_id^\dagger_{\alpha,s}c_{\gamma,s}P_j}{\epsilon_0+U({i+j-1)}-\epsilon_{\gamma}}-{\rm H.c.}.
\end{equation}
Proof: given that $[P_i,c_{\gamma,s}]=0$, we have
\begin{eqnarray}
[H_0,S]&=&\sum_{ij,\alpha,\gamma,s}V^\alpha_{\gamma}\frac{[H_0,P_id^\dagger_{\alpha,s}c_{\gamma,s}P_j]}{\epsilon_0+U(i+j-1)-\epsilon_{\gamma}}+{\rm H.c.}\nonumber\\
&=&\sum_{ij,\alpha,\gamma,s}V^\alpha_{\gamma}\frac{[H_0,P_id^\dagger_{\alpha,s}P_j]c_{\gamma,s}}{\epsilon_0+U(i+j-1)-\epsilon_{\gamma}}\nonumber\\
&+&\sum_{ij,\alpha,\gamma,s}V^\alpha_{\gamma}\frac{P_id^\dagger_{\alpha,s}P_j[H_0,c_{\gamma,s}]}{\epsilon_0+U(i+j-1)-\epsilon_{\gamma}}+{\rm H.c.}.\nonumber
\end{eqnarray}
We then have that
\begin{eqnarray}
[H_0,P_id^\dag_{\alpha,s}P_j]&=&P_i\left[\epsilon_0 d^\dagger_{\alpha,s}+U(\hat{N}d^\dagger_{\alpha,s}+d^\dagger_{\alpha,s}\hat{N}{\color{red}-} d^\dagger_{\alpha,s})\right]P_j\nonumber\\
&=&P_i\left[\epsilon_0 d^\dagger_{\alpha,s}+U(i+j-1)d^\dagger_{\alpha,s}\right]P_j \nonumber\\
&=&\left(\epsilon_0+U(i+j-1)\right)P_id^\dagger_{\alpha,s}P_j \nonumber
\end{eqnarray}
and
\begin{eqnarray}
[P_id_{\alpha,s}P_j,H_0] &=&P_i\left[\epsilon_0 d_{\alpha,s}+U(\hat{N}d_{\alpha,s}+d_{\alpha,s}\hat{N}{\color{red}-} d_{\alpha,s})\right]P_j\nonumber\\
&=&P_i\left[\epsilon_0  d_{\alpha,s}+U(i+j-1)d_{\alpha,s}\right]P_j \nonumber\\
&=&\left(\epsilon_0+U(i+j-1)\right)P_id_{\alpha,s}P_j \nonumber\\
\end{eqnarray}
where we used $P_j\hat{N}=\hat{N}P_j=jP_j$ being $[\hat{N}(\hat{N}-1),d^\dagger_{\alpha,s}]= \hat{N}[\hat{N},d^\dagger_{\alpha,s}]+[\hat{N},d^\dagger_{\alpha,s}](\hat{N}-1)$ and $[\hat{N}(\hat{N}-1),d_{\alpha,s}]=-\hat{N}[\hat{N},d_{\alpha,s}]-[\hat{N},d_{\alpha,s}](\hat{N}-1)$. 
It follows that
\begin{eqnarray}
[H_0,S]
&=&\sum_{ij,\alpha,\gamma,s}V^\alpha_{\gamma}\frac{(\epsilon_0+(i+j-1)U)P_id^\dag_{\alpha,s}P_jc_{\gamma,s}}{\epsilon_0+U(i+j-1)-\epsilon_{\gamma}}\nonumber\\
&-&\sum_{ij,\alpha,\gamma,s}V^\alpha_{\gamma}\frac{\epsilon_{\gamma,s}P_id^\dag_{\alpha,s}P_jc_{\gamma,s}}{\epsilon_0+U(i+j-1)-\epsilon_{\gamma}}+{\rm H.c.}\nonumber\\
&=&\sum_{ij,\alpha,\gamma,s}V^\alpha_{\gamma}P_id^\dag_{\alpha,s}P_jc_{\gamma,s}+{\rm H.c.}\nonumber\\
&=&\sum_{\alpha,\gamma,s}V^\alpha_{\gamma}d^\dag_{\alpha,s}c_{\gamma,s}+{\rm H.c.}\nonumber
\end{eqnarray}
where in the last equation we used $\sum_iP_i=1$. The effective Hamiltonian now takes the form $H_{\rm eff}=H_0+H'$, with
\begin{equation}\label{H'}
H'=\frac{1}{2}[S,V].
\end{equation}

\subsection{Effective Hamiltonian in the $P_2$ subspace}

Out of the four possible combinations arising from the commutators we neglect terms like $d^\dag d^\dag c c$ and $ddc^\dagger c^\dagger$. We then have
\begin{eqnarray}
H'&=&\frac{1}{2}\sum_{ij,\alpha,\mu,s}V^\alpha_{\mu}(V^\beta_\nu)^*\frac{[P_id^\dag_{\alpha,s}c_{\mu,s}P_j,c^\dag_{\nu,s'}d_{\beta,s'}]}{\epsilon_0+U(i+j-1)-\epsilon_{\mu}}\nonumber\\
&-&\frac{1}{2}\sum_{ij,\alpha,\mu,s}(V^\alpha_{\mu})^*V^\beta_\nu\frac{[P_jc^\dag_{\mu,s}d_{\alpha,s}P_j,d^\dag_{\beta,s'}c_{\nu,s'}]}{\epsilon_0+U(i+j-1)-\epsilon_{\mu}}
\end{eqnarray}
We notice that
\begin{eqnarray}
[P_id^\dag_{\alpha,s}c_{\mu,s}P_j,c^\dag_{\nu,s'}d_{\beta,s'}]&=&P_id^\dag_{\alpha,s}P_jd_{\beta,s}\delta_{\mu,\nu}\delta_{s,s'}\nonumber\\
&-&c^\dag_{\nu,s'}c_{\mu,s}\left\{P_id^\dag_{\alpha,s}P_j,d_{\beta,s'}\right\}\nonumber\\
\end{eqnarray}
where $\{A,B\}=AB+BA$. The first term renormalizes the energy of the dots and the second term gives us the desired interaction term, 
\begin{eqnarray}
H'&=&-\frac{1}{2}\sum_{ij\alpha\mu s,\beta\nu s'}V^\alpha_\mu(V^\beta_\nu)^*c^\dag_{\nu s'}c_{\mu s}\nonumber\\
&\times&\left[\frac{\{P_id^\dag_{\alpha s}P_j,d_{\beta s'}\}}{\epsilon_0+U(i+j-1)-\epsilon_\mu}+\frac{\{P_jd_{\beta s'}P_i,d^\dag_{\alpha s}\}}{\epsilon_0+U(i+j-1)-\epsilon_\nu}\right]
\nonumber\\
\end{eqnarray}
The term $P_id^\dag_{\alpha,s}P_j$ and $P_jd_{\beta,s'}P_i$ can be non-zero only if $i=j+1$. Due to the fact that our ground state is described by a 2 electrons occupancy, we are going to focus in that case. In other words, we project the effective Hamiltonian on the subspace with occupancy 2, with one electron per dot. This is done in two steps,
\begin{equation}
H'\to P_2H'P_2\to  P_{1L}P_{1R} H'P_{1L}P_{1R}, 
\end{equation}
The first step selects intermediate states with $j=1,2$, so that we have
\begin{eqnarray}
H'&=&-\frac{1}{2}\sum_{\alpha\mu s,\beta\nu s'}A^{(1)}_{\alpha\mu,\beta\nu}c^\dag_{\nu s'}c_{\mu s}P_{2}d^\dag_{\alpha s}P_1,d_{\beta s'}P_2\nonumber\\
&-&\frac{1}{2}\sum_{\alpha\mu s,\beta\nu s'}A^{(2)}_{\alpha\mu,\beta\nu}c^\dag_{\nu s'}c_{\mu s}P_{2}d_{\beta s'}P_3d^\dag_{\alpha s}P_2,
\end{eqnarray}
where
\begin{equation}
A^{(j)}_{\alpha\mu,\beta\nu}=V^\alpha_\mu(V^\beta_\nu)^*\left[\frac{1}{\epsilon_0+2Uj-\epsilon_\mu}+\frac{1}{\epsilon_0+2Uj-\epsilon_\nu}\right],
\end{equation}

In the second step we operate with $P_{1L}P_{1R}$ from the left and the right of $H'$. The following relations hold
\begin{eqnarray}
P_{L1}P_{R1}d^\dag_{\alpha s}P_1,d_{\beta s'}P_{L1}P_{R1}&=&\delta_{\alpha\beta}P_{1\bar{\alpha}}d^\dag_{\alpha s}d_{\alpha s'}\\
P_{L1}P_{R1}d_{\beta s'}P_3,d^\dag_{\alpha s}P_{L1}P_{R1}&=&\delta_{\alpha\beta}P_{1\bar{\alpha}}P_{1\alpha}d_{\alpha s'}P_{2\alpha}d^\dag_{\alpha s}P_{1\alpha}\nonumber\\
&=&\delta_{\alpha\beta}P_{1\bar{\alpha}}(1-d^\dag_{\alpha s}d_{\alpha s'})
\end{eqnarray}
The second expression generates a term that renormalizes the graphene's electron energy. Collecting the different terms we find
\begin{eqnarray}
H'&=&-\frac{1}{2}\sum_{\alpha\mu s,\nu s'}(A^{(1)}_{\alpha\mu,\alpha\nu}-A^{(2)}_{\alpha\mu,\alpha\nu})c^\dag_{\nu s'}c_{\mu s}d^\dag_{\alpha s}d_{\alpha s'}\nonumber\\
&-&\frac{1}{2}\sum_{\alpha\mu s,\nu s'}A^{(2)}_{\alpha\mu,\alpha\nu}c^\dag_{\nu s'}c_{\mu s}
\end{eqnarray}
where we set $P_{1\bar{\alpha}}=1$.
We can reorder the expression as
\begin{equation}
H'=H_{\rm K}+H_{ch},
\end{equation}
where the Kondo term is given by
\begin{eqnarray}
H_{\rm K}&=&-\sum_{\mu\mu'} \frac{J^\alpha_{\mu,\mu'}}{2}\left[c^\dag_{\mu'\downarrow}c_{\mu\uparrow}d^\dag_{\alpha\downarrow}d_{\alpha \uparrow}+c^\dag_{\mu'\uparrow}c_{\mu\downarrow}d^\dag_{\alpha\uparrow}d_{\alpha \downarrow}\right.\nonumber\\
&+&\frac{1}{2}\left.(c^\dag_{\mu'\uparrow}c_{\mu \uparrow}-c^\dag_{\mu'\downarrow}c_{\mu \downarrow})(d^\dag_{\alpha\uparrow}d_{\alpha \uparrow}-d^\dag_{\alpha\downarrow}d_{\alpha \downarrow})\right],
\end{eqnarray}
with $J^\alpha_{\mu\mu'}=A^{(1)}_{\alpha\mu,\alpha\mu'}-A^{(2)}_{\alpha\mu,\alpha\mu'}$ and the charge term $H_{ch}$ reads
\begin{equation}
H_{ch}=-\sum_{\alpha,\mu\mu',ss'} c^\dag_{\mu's}c_{\mu s}\left[W^\alpha_{\mu,\mu'}+ \frac{J^\alpha_{\mu,\mu'}}{4}d^\dag_{\alpha s'}d_{\alpha s'}\right].
\end{equation}
with $W^\alpha_{\mu\mu'}=A^{(2)}_{\alpha\mu,\alpha\mu'}/2$.

Analogously, it is possible to derive the effective Hamiltonian in the $P_{2L}P_{0R}+P_{0L}P_{2R}$ subspace, staring from the expression Eq.~(\ref{H'}) and performing the following steps
\begin{equation}
H'\to P_2H'P_2 \to  (P_{2L}P_{0R}+P_{0L}P_{2R}) H' P_{1L}P_{1R}
\end{equation}

\subsection{Channel decoupling}

Neglecting the momentum dependence of the coupling $J$, the Kondo Hamiltonian acquires a particularly simple form
\begin{eqnarray}
H_{\rm K}&=&-\frac{J}{2}\sum_{{\bf k},{\bf k}',\alpha,s} C^\dag_{\alpha{\bf k}'\bar{s}}C_{\alpha{\bf k}s}d^\dag_{\alpha s}d_{\alpha \bar{s}}\nonumber\\
&-&\frac{J}{4}\sum_{{\bf k},{\bf k}',\alpha,s,s'} ss'C^\dag_{\alpha,{\bf k},s}C_{\alpha,{\bf k}',s}d^\dag_{\alpha s'}d_{\alpha s'},
\end{eqnarray}
In the base of the graphene sublattice operators $a,b$, is the two independent Kondo terms are given by
\begin{eqnarray}
H^L_K&=&-\frac{J}{8}\sum_{{\bf k},{\bf k}',ss'}ss'(a^\dagger_{K,s}+b^\dagger_{K',s})(a_{K,s}+b_{K',s})n_{L,s'}\nonumber\\
&-&\frac{J}{4}\sum_{{\bf k},{\bf k}',\alpha,s}(a^\dagger_{K,\bar{s}}+b^\dagger_{K',\bar{s}})(a_{K,s}+b_{K',s})d^\dagger_{L,s}d_{L,\bar{s}},~~~~~~\\
H^R_K&=&-\frac{J}{8}\sum_{{\bf k},{\bf k}',\alpha,ss'}ss'(a^\dagger_{K',s}+b^\dagger_{K,s})(a_{K',s}+b_{K,s})n_{R,s'}\nonumber\\
&-&\frac{J}{4}\sum_{{\bf k},{\bf k}',\alpha,s}(a^\dagger_{K',\bar{s}}+b^\dagger_{K,\bar{s}})(a_{K',s}+b_{K,s})d^\dagger_{R,s}d_{R,\bar{s}}.~~~~~~
\end{eqnarray}
We can see how each orbital (L and R) is coupled with a different channel in graphene. 

\end{document}